\pdfoutput=1
\documentclass[conference]{IEEEtran}

\usepackage{graphicx}
\usepackage{tabularx}

\usepackage[T1]{fontenc}
\usepackage[utf8]{inputenc}

\usepackage{csquotes}
\usepackage{nicefrac}
\usepackage[textsize=scriptsize]{todonotes}
\usepackage{booktabs}
\usepackage{xcolor}
\usepackage{tikz}
\usetikzlibrary{arrows,positioning} 
\usetikzlibrary{shapes.geometric, positioning, quantikz}
\usepackage{nicematrix}
\usepackage{flushend}
\usepackage{hyperref}
\usepackage[binary-units=true, detect-all=true]{siunitx}
\usepackage{etoolbox}
\usepackage{soul}
\usepackage{amssymb}
\usepackage{booktabs} %
\usepackage{csvsimple}
\usepackage{amsthm}

\robustify\bfseries

\makeatletter
\let\MYcaption\@makecaption
\makeatother
\usepackage[font=footnotesize]{subcaption}
\makeatletter
\let\@makecaption\MYcaption
\makeatother

\usepackage{yquant}
\usepackage{pgfplots}

\usepackage[style=ieee, maxnames=4, minnames=1, date=year, doi=false,isbn=false,backend=biber]{biblatex}
\usepackage[binary-units=true, detect-all=true]{siunitx}
\usepackage{etoolbox}
\robustify\bfseries

\usepackage{nicematrix}
\usetikzlibrary{fit, quotes}
\usepackage{bbm}

\newtheorem{example}{Example}

\usetikzlibrary{automata,arrows}

\AtEveryBibitem{
  \clearname{editor}%
  \clearfield{series}%
  \clearfield{isbn}%
  \clearfield{issn}%
  \clearfield{volume}%
  \clearfield{number}%
  \clearfield{pages}%
  \clearfield{doi}%
}

\begin{document}

\title{Compiler Optimization for Quantum Computing \\ Using Reinforcement Learning}

\author{
	\IEEEauthorblockN{Nils Quetschlich\IEEEauthorrefmark{1}\hspace*{1.5cm}Lukas Burgholzer\IEEEauthorrefmark{2}\hspace*{1.5cm}Robert Wille\IEEEauthorrefmark{1}\IEEEauthorrefmark{3}}
	\IEEEauthorblockA{\IEEEauthorrefmark{1}Chair for Design Automation, Technical University of Munich, Germany}
	\IEEEauthorblockA{\IEEEauthorrefmark{2}Institute for Integrated Circuits, Johannes Kepler University Linz, Austria}
	\IEEEauthorblockA{\IEEEauthorrefmark{3}Software Competence Center Hagenberg GmbH (SCCH), Austria}
	\IEEEauthorblockA{\href{mailto:nils.quetschlich@tum.de}{nils.quetschlich@tum.de}\hspace{1.5cm}\href{mailto:lukas.burgholzer@jku.at}{lukas.burgholzer@jku.at}\hspace{1.5cm} \href{mailto:robert.wille@tum.de}{robert.wille@tum.de}\\
	\url{https://www.cda.cit.tum.de/research/quantum}
	}
}

\maketitle

\begin{abstract}
Any quantum computing application, once encoded as a quantum circuit, must be compiled before being executable on a quantum computer.
Similar to classical compilation, quantum compilation is a sequential process with many compilation steps and numerous possible optimization passes.
Despite the similarities, the development of compilers for quantum computing is still in its infancy---lacking mutual consolidation on the best sequence of passes, compatibility, adaptability, and flexibility.
In this work, we take advantage of decades of classical compiler optimization and propose a \emph{reinforcement learning} framework for developing optimized quantum circuit compilation flows. 
Through distinct constraints and a unifying interface, the framework supports the combination of techniques from different compilers and optimization tools in a single compilation flow.
Experimental evaluations show that the proposed framework---set up with a selection of compilation passes from IBM's Qiskit and Quantinuum's TKET---significantly outperforms both individual compilers in $73\%$ of cases regarding the expected fidelity.
The framework is available on GitHub (\url{https://github.com/cda-tum/MQTPredictor}) as part of the Munich Quantum Toolkit (MQT).
\end{abstract}

\section{Introduction}
\label{sec:intro}

In classical computing, each \emph{Central Processing Unit} (CPU) comes with its own \emph{Instruction Set Architecture} (ISA)---describing the native operations the CPU can process.
\mbox{High-level} programming languages such as \emph{C++}, \emph{Rust}, or \emph{Go} are used to develop programs in a \mbox{platform-agnostic} way.
These high-level descriptions are then \emph{compiled} to a specific device's ISA by a \emph{compiler}.
Prominent examples are \emph{LLVM}~\cite{lattner2004llvm} or \emph{GCC} for C++, \emph{rustc} for Rust, and \emph{gc} or \emph{gccgo} for Go.
In each of these compilers, a plethora of passes for analyzing and transforming the source code is available.
The \enquote{quality} of the code generated by compilers heavily depends on the selection of the individual passes applied during the compilation process as well as their execution order.
Furthermore, one particular combination is highly unlikely to be the most effective sequence for every program on a particular machine.
Even nowadays, after decades of classical compiler research, determining good sequences of compilation passes turns out to be a challenging task.
In the literature, this problem is commonly referred to as \emph{Phase Ordering}.

Research on the phase ordering problem already started roughly four decades ago~\cite{phase1, phase2, codetransformationsWhitfield1997} and is still very actively pursued.
In the recent years, \emph{autotuning}~\cite{autotuning_compiler} (trying out different compilation parameters and evaluating the results with respect to a particular metric) and machine learning~\cite{ml_in_compilers} have shown promising results.
A particularly promising branch of research is the application of \emph{Reinforcement Learning}~(RL,~\cite{kaelbling1996reinforcement, sutton2018reinforcement}) for this purpose.
In~\cite{reinforcement2}, the authors show that their \mbox{RL-based} approach to phase ordering outperforms LLVM's optimization scheme.
Reinforcement learning has also successfully been applied for inlining optimization~\cite{reinforcement1} and vectorization~\cite{reinforcement3}.

The compilation of quantum circuits/programs follows a very similar scheme as the compilation of classical programs.
All major quantum SDKs (such as IBM's Qiskit~\cite{qiskit}, Quantinuum's TKET~\cite{tket}, AWS Braket~\cite{AmazonBraketPython2022}, or Microsoft's Azure Quantum~\cite{bradbenAzureQuantumDocumentation}) implement some sort of compilation flow for making circuits executable on particular devices (see, e.g.,~\cite{wille2019ibmtoolchain}).
However, while research on classical compilers has been conducted for decades, compilers for quantum computing are still in early development.
This manifests in several ways:
\begin{enumerate}
	\item There is no mutual consolidation on what the best methods and combinations thereof are.
	\item It is not straightforward to mix and match different options from various frameworks---creating something similar to a vendor \mbox{lock-in}.
	\item Major SDKs can be quite slow in adopting the newest research findings---effectively restraining users from taking advantage of state-of-the-art methods.
	\item Users can hardly customize the optimization goal or the target metric for the compilation process. Instead, they have to rely on predefined combinations of compilation passes provided by the SDKs---typically in the form of various optimization levels.
\end{enumerate}
As a consequence, we are quickly heading towards the same problems as in classical compiler development and, despite the similarities, techniques from the classical realm cannot be straightforwardly applied in the quantum realm due to the inherent differences of both computational paradigms.

In this work, we aim to mitigate the drawbacks discussed above.
Instead of trying to reinvent the wheel, the solution proposed in this work builds upon the decades of research on compiler optimization in the classical realm.
To this end, we adapt an established classical technique to the quantum domain---modeling the compilation flow as a \emph{Markov Decision Process} (MDP,~\cite{mdp}) and applying \emph{Reinforcement Learning} to learn the best order in which the different compilation passes and optimizations can be applied.

The resulting framework (which is available on GitHub (\url{https://github.com/cda-tum/MQTPredictor}) as part of the Munich Quantum Toolkit (MQT)) is platform agnostic and easily extendable with further compilation methods from different SDKs and software tools.
This is accomplished by defining distinct constraints that can be efficiently checked for each state in the process and a unifying interface for each compilation step.
In contrast to slowly-adapting compilers in major SDKs, this allows to develop optimized compilation flows that can keep up with the fast-paced development and research in this domain.

Experimental evaluations demonstrate the feasibility of the proposed approach.
Compared to the compilation flows offered by IBM's Qiskit and Quantinuum's TKET, the proposed framework---set up with a selection of compilation passes from both tools---is shown to outperform both individual compilers with respect to expected fidelity, critical depth, and their combination in $73\%$, $84\%$, $75\%$ of cases.

The rest of this paper is structured as follows.
\autoref{sec:background} briefly reviews quantum circuit compilation.
Then, \autoref{sec:proposed} describes how compiler optimization for quantum circuits can be modeled as an MDP and how it can be tackled with reinforcement learning.
Based on that, \autoref{sec:exp} provides details on the implementation of the proposed framework and evaluates its performance.
Finally, \autoref{sec:conclusions} concludes the paper.

\vspace{-1mm}
\section{Background: Quantum Circuit Compilation}\label{sec:background}
\vspace{-1mm}

Quantum programs are commonly described in the form of \emph{quantum circuits}, i.e., sequences of quantum operations (also called quantum gates) acting on the qubits of a quantum system.

\begin{figure}[t]
\centering
     \begin{subfigure}{0.49\linewidth}    
\centering 

				\begin{tikzpicture}
				  \begin{yquant}
				    	qubit {${q_2}$} q;    	
						qubit {${q_1}$} q[+1];
						qubit {${q_0}$} q[+1];
						
						[name=left]
				    	box {$X$} q[0];
				    	[name=right]
				    	box {$X$} q[0];
				    	
				    	box {$H$} q[1];
				    	box {$R_z(-\frac{\pi}{2})$} q[1];
				    	
				    	cnot q[1] | q[0];
				    	cnot q[2] | q[1];
				    	cnot q[2] | q[0];
				    	
				  \end{yquant}
				  \node[fit=(left)(right), draw, green, rounded corners, inner sep=2pt] {};
				\end{tikzpicture}
		\caption{Quantum circuit.}
         \label{fig:sub_1}
		 \end{subfigure}\vspace{2mm}\hfill\\%
\begin{subfigure}{\linewidth}
\centering
				\begin{tikzpicture}
				  \begin{yquant}
				    	qubit {${q_2}$} q;    	
						qubit {${q_1}$} q[+1];
						qubit {${q_0}$} q[+1];

						[name=left_decomp]
				    	box {$R_z(\frac{\pi}{2})$} q[1];
				    	box {$\sqrt{X}$} q[1];
						[name=left]
				    	box {$R_z(\frac{\pi}{2})$} q[1];
				    	[name=right]
				    	box {$R_z(-\frac{\pi}{2})$} q[1];
				    	
				    	cnot q[1] | q[0];
				    	cnot q[2] | q[1];
				    	cnot q[2] | q[0];
				    	
				  \end{yquant}
				  \node[fit=(left)(right), draw, green, rounded corners, inner sep=2pt] {};
				  \node[fit=(left_decomp)(left), draw, orange, inner ysep=3pt, inner xsep=2pt, rounded corners] {};
				\end{tikzpicture}
         \caption{Compilation to only native gates.}
         \label{fig:sub_2}
     \end{subfigure}\vspace{2mm}\hfill\\%
          		 \begin{subfigure}[b]{\linewidth}
					\centering		
					   \begin{tikzpicture}
						    \node (p0) at ( 0, 1) [circle, draw]{$Q_0$}; 
						    \node (p1) at ( 1,1) [circle, draw]{$Q_1$};
						    \node (p2) at ( 2,1) [circle, draw]{$Q_2$};
						    \begin{scope}[every path/.style={-}]
						       \draw (p0) -- (p1);
						       \draw (p1) -- (p2);
						    \end{scope}  
						\end{tikzpicture}	

			         \caption{Target device with three physical qubits.}
			         \label{fig:arch}
		     \end{subfigure}\vspace{2mm}\hfill\\%
     \begin{subfigure}[b]{\linewidth}
\centering
\resizebox{0.9\linewidth}{!}{
		\begin{tikzpicture}
  \begin{yquant}
    	qubit {${q_2} \mapsto Q_2$} q;    	
		qubit {${q_1} \mapsto Q_1$} q[+1];
		qubit {${q_0} \mapsto Q_0$} q[+1];

		box {$R_z(\frac{\pi}{2})$} q[1];
		box {$\sqrt{X}$} q[1];

	    cnot q[1] | q[0];
		[name=left]
	    
		cnot q[2] | q[1];
		[name=right]
		cnot q[2] | q[1];
		cnot q[1] | q[2];
		[name=right_decomp]
		cnot q[2] | q[1];
		cnot q[1] | q[0];

  \end{yquant}
\node[fit=(left)(right), draw, green, inner ysep=10pt, yshift=0.28cm, rounded corners] {};
				  \node[fit=(right)(right_decomp), draw, orange, yshift=0.28cm, inner ysep=12pt, rounded corners] {};
\end{tikzpicture}}
         \caption{Compilation to target architecture.}
         \label{fig:sub_3}\vspace*{-3mm}
     \end{subfigure}
        \caption{Compilation of a quantum circuit to a targeted device.}
        \label{fig:compflow}\vspace*{-5mm}
\end{figure}
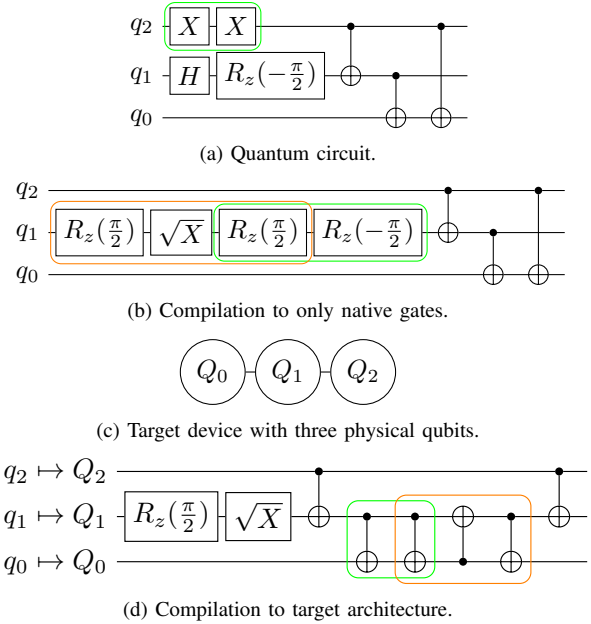

\begin{example}\label{ex:compflow1}
A simple quantum circuit operating on three qubits ($q_0$, $q_1$, $q_2$) is shown in \autoref{fig:sub_1}.
It contains four different kinds of gates: Two NOT gates (denoted as \emph{X}), one Hadamard gate (denoted as \emph{H}), one $Z$-rotation gate (denoted as \emph{$R_z$}), and three two-qubit \mbox{controlled-NOT} or CNOT gates (with their control and target qubits denoted as $\bullet$ and $\oplus$, respectively).
\end{example}

Executing such programs on an actual \emph{Quantum Processing Unit} (QPU) requires \emph{compilation} conducted by \emph{compilers}.
They usually first employ device-independent optimizations to reduce the complexity of the original circuit description.
Examples of such optimizations include \mbox{gate-cancellation}, \mbox{gate-commutation}, or \mbox{high-level} synthesis routines, e.g., proposed in ~\cite{hattoriQuantumCircuitOptimization2018, niemann2020advancedexactsynt, vidalUniversalQuantumCircuit2004, itokoQuantumCircuitCompilers2019, maslovQuantumCircuitSimplification2008}.

\begin{example}\label{ex:compflow2}
	Consider again the circuit shown in \autoref{fig:sub_1}. 
	Employing \mbox{device-independent} optimizations would, e.g., lead to the cancellation of the consecutive $X$ gates on $q_2$ (highlighted by the green box) since $X\cdot X = I$.
\end{example}

After these initial optimizations, \mbox{QPU-specific} circuits are generated.
To this end, each quantum processor offers a particular set of \emph{native gates}, i.e., operations that can be directly executed on the device.
Consequently, all \mbox{non-native} gates need to be \emph{synthesized} using these native gates for the circuit to be executable.
This synthesis step creates opportunities for further optimizations.

\begin{example}\label{ex:compflow3}
Recent devices offered by IBM provide a native gate-set consisting of \emph{$R_z$}, \emph{$\sqrt{X}$}, \emph{X}, and \emph{CNOT} gates.
To execute the circuit from \autoref{fig:sub_1} on such a device, e.g., the Hadamard gate needs to be expressed in terms of these gates.
A suitable decomposition into two $R_z(\frac{\pi}{2})$ and a $\sqrt{X}$ gate is shown in \autoref{fig:sub_2} (highlighted by the orange box).
Incidentally, one of the $R_z(\frac{\pi}{2})$ gates can, then, be cancelled with the existing $R_z(-\frac{\pi}{2})$ gate (highlighted by the green box).
\end{example}

Some QPUs (e.g., those based on superconducting qubits or neutral atoms) only feature a limited connectivity between their qubits, so that \mbox{multi-qubit} gates can only be applied to qubits that are connected on the device.
In these cases, a \emph{mapping} step is required. 
Mapping is frequently split into \emph{layouting}, i.e., assigning the circuit's logical qubits to the device's physical qubits, and \emph{routing}, i.e., ensuring the connectivity constraints are satisfied by dynamically changing the assignment of logical to physical qubits.
Again, optimizations may be applied after mapping the circuit to reduce the induced overhead.

\begin{example}\label{ex:compflow4}
Assume the circuit from \autoref{fig:sub_2} shall be executed on an IBM device with three physical qubits ($Q_0$, $Q_1$, $Q_2$) arranged in a line (as depicted in \autoref{fig:arch}), i.e., only interactions between $Q_0$ and $Q_1$ as well as $Q_1$ and $Q_2$ are possible.
Then, no layout exists that makes this circuit executable right away since the circuit contains interactions between all pairs of qubits.
Thus, a SWAP gate (eventually realized as a sequence of three CNOT gates highlighted by the orange box) needs to be inserted during the routing step to make the circuit executable (as shown in \autoref{fig:sub_3}).
In this case, some of the induced overhead can be reduced by cancelling two CNOT gates (highlighted by the green box).
\end{example}

\begin{figure*}
\centering
\resizebox{.8\linewidth}{!}{
\begin{tikzpicture}[auto]
\def\w{1.5pt}

\node (p0) at ( 0, 1) [circle, draw, minimum size=1.8cm]{Start};
    \node (p1) at ( 5,1) [circle, draw, align=center, minimum size=1.8cm]{Platform \\chosen};
    \node (p2) at ( 10,1) [circle, draw, align=center, minimum size=1.8cm]{Device \\chosen};
    \node (p3) at ( 15,1) [circle, draw, align=center, minimum size=1.8cm]{Only \\native \\gates};
    \node (p4) at ( 20,1) [accepting, circle, draw, minimum size=1.8cm,thick]{Done};

    \path[->] (p0) edge [line width=\w] node [align=center]  {\emph{Quantum platform} \\ \emph{selection}} (p1);
    \path[->] (p1) edge [line width=\w] node [align=center]  {\emph{Quantum device} \\ \emph{selection}} (p2);
    \path[->] (p2) edge [line width=\w] node [align=center]  {\emph{Synthesis}} (p3);
    \path[->] (p3) edge [line width=\w] node [align=center]  {\emph{Mapping*}} (p4);
                                                
    \def\lin{110}
    \def\lout{70}
    \def\lloose{5}
    \tikzset{every loop/.style={in=\lin,out=\lout,looseness=\lloose}}                                              
    \path[->]  (p0) edge [loop above, color=blue, line width=\w] node [align=center]{}(p0);
    \path[->]  (p2) edge [loop above, color=blue, line width=\w] node [align=center]{}(p2);
    \path[->]  (p3) edge [loop above, color=blue, line width=\w] node [align=center]{}(p3);
    \path[->]  (p4) edge [loop above, color=blue, line width=\w] node [align=center]{}(p4);

\def\h{0.4}
\draw[->, blue, line width=\w] (p4.265) .. controls ($(p4.south) - (0.1,\h)$) .. ($(p4.south) - (1, \h)$) -- ($(p3.south) - (-1.5, \h)$) .. controls ($(p3.south) - (-1.2, \h)$) .. (p3.315);
\draw[->, blue, line width=\w] (p3.270) .. controls ($(p3.south) - (0,\h)$) .. ($(p3.south) - (1, \h)$) -- ($(p2.south) - (-1.5, \h)$) .. controls ($(p2.south) - (-1.2, \h)$) .. (p2.315);
\def\hd{0.55}
\draw[->, blue, line width=\w] (p4.275) .. controls ($(p4.south) - (-0.1,\hd)$) .. ($(p4.south) - (1, \hd)$) -- ($(p2.south) - (-1.2, \hd)$) .. controls ($(p2.south) - (0, \hd)$) .. (p2.270);
\draw[->, line width=\w] ($(p0.west) - (0.5, 0)$) -- (p0.west);

    \node (opt) [text=blue, below left=0.9cm and -1cm  of  p4] {Optimization};
\node [text=black, below left=1.4cm and -1cm  of  p4] {*might not be necessary for all platforms};
\end{tikzpicture}}
\vspace{-3mm}
\caption{Quantum compilation flow modeled as an MDP.}
\label{fig:mdp}
\vspace{-5mm}
\end{figure*}
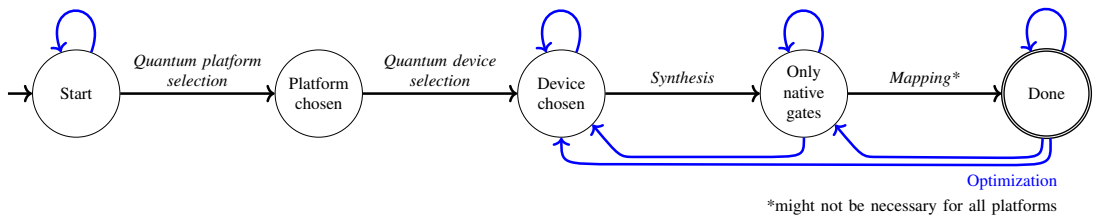

\vspace{5cm}

With the number of different synthesis, mapping, and optimization techniques for quantum circuits rapidly increasing, the landscape of compiler passes is becoming more and more complex---leading to several obstacles as discussed in \autoref{sec:intro}.
In this work, we propose a framework that aims to overcome these obstacles.

\section{Proposed Framework}\label{sec:proposed}

Fortunately, the underlying problem is not new per-se, since classical compilers have seen a similar trend in the past and lots of research has been conducted to find efficient solutions there.
In the following, we describe how a classically-inspired reinforcement learning technique can be adapted to quantum circuit compilation.
To this end, we first describe how to programmatically model quantum circuit compilation in order to, then, apply reinforcement learning to try to learn the best sequence of compilation passes depending on characteristics of the initial circuit---resulting in an \emph{optimized compiler} for quantum circuits.

\subsection{Modeling as a Markov Decision Process}

The quantum circuit compilation flow reviewed above can be seen as a sequential procedure that takes an initial (\mbox{high-level}) quantum circuit and, step by step, applies certain actions/transformations to eventually arrive at a circuit representation that can be executed on a target device.
Such a procedure can be formally modeled as a \emph{Markov Decision Process} (MDP,~\cite{mdp}).
To this end, it is split into states and corresponding actions that can be performed in each of the states.
The model proposed in this work is illustrated in \autoref{fig:mdp}---with states being denoted as circles and actions as arrows.
In the following, the process with its respective actions is described one by one.

Initially, the process starts in the \enquote{Start} state and stores the original quantum circuit to be compiled.
At this point, there are two possible types of actions.
On the one hand, \mbox{device-independent} \emph{optimizations} can be applied to the circuit (indicated by the blue arrow).
As reviewed in \autoref{sec:background}, this includes techniques such as single-qubit gate fusion, gate cancellation and commutation, as well as high-level synthesis methods, e.g., proposed in~\cite{hattoriQuantumCircuitOptimization2018, niemann2020advancedexactsynt, vidalUniversalQuantumCircuit2004, itokoQuantumCircuitCompilers2019, maslovQuantumCircuitSimplification2008}.
After such an optimization, the process remains in the same state as before, but now stores a new optimized quantum circuit.
On the other hand, the process can proceed to a different state by \emph{Quantum platform selection} actions, e.g., choosing \emph{IBM}, \emph{Rigetti}, or \emph{IonQ}---effectively fixing the targeted native gate-set.
In the classical domain, this roughly corresponds to deciding, e.g., between \emph{x86} and \emph{ARM} architectures.

Once a platform has been chosen (i.e., the process is in the \enquote{Platform chosen} state), only a single type of action can be applied.
Namely, the \emph{Quantum device selection} actions offered by the selected platform.
This step fixes the number of qubits and their topology, i.e., which qubits can interact with each other.
Depending on the choice of the platform, a varying range of devices with different characteristics is available.
At the time of writing, IBM for example offers 23 different quantum computers that each vary in their qubit count, their quantum volume, and their \emph{Circuit Layer Operations Per Second} (CLOPS).
In the classical domain, this roughly corresponds to targeting a specific CPU model family, such as \emph{Intel Alder Lake} or \emph{AMD Zen 4}.

After a particular device has been selected (i.e., the process is in the \enquote{Device chosen} state), the \mbox{device-dependent} compilation begins.
The goal of the remaining steps is to produce a circuit that can be executed on the chosen device.
To this end, the circuit needs to satisfy two conditions:
\begin{enumerate}
	\item it must only use gates native on the selected platform, and
	\item it must conform to the topology of the selected device.
\end{enumerate}
While the first condition can be ensured by using \emph{Synthesis} methods, e.g., as proposed in~\cite{gilesExactSynthesisMultiqubit2013, amyMeetinthemiddleAlgorithmFast2013, millerElementaryQuantumGate2011, schneiderSATEncodingOptimal2023}, the second condition is realized by corresponding \emph{Mapping} methods, e.g., as proposed in~\cite{siraichiQubitAllocation2018, zulehnerEfficientMethodologyMapping2019, matsuoEfficientMethodQuantum2019}. 
Both of these tasks are highly \mbox{non-trivial} and there is a vast variety of methods available in major quantum SDKs as well as individual research tools.
At the time of writing, IBM's Qiskit alone offers four different layout and five routing algorithms---leading to 20 different mapping schemes.
Once both conditions are satisfied, the process is in the \enquote{Done} state, i.e., the circuit is executable.
Optimizations may be applied in any state during the \mbox{device-dependent} compilation.
Depending on whether the resulting circuit satisfies none, the native-gates, or both constraints, the process respectively continues in the \enquote{Device chosen}, \enquote{Only native gates}, or \enquote{Done} state (again, indicated by the blue arrows).

Overall, the formulation described above leads to a modular framework with two characteristic traits:
\begin{enumerate}
	\item Each state has \emph{distinct constraints} that can be efficiently checked, e.g., whether the circuit only contains certain native gates or only acts on qubits connected on the device. Thus, it is \mbox{straight-forward} to determine the state after a certain action has been applied.
	\item All actions have a \emph{unified interface}, i.e., all of them use a quantum circuit as the main representation for their input and output. Moreover, the same format is used independent of the SDK or tool the actions originate from.
\end{enumerate}
As a consequence, it is possible to mix and match compiler passes from different SDKs and tools and integrate them into a single \emph{optimized} compiler.
Such a framework also allows to mitigate vendor \mbox{lock-ins} and can help to keep up with the rapid pace of the field by allowing one to quickly integrate new methods into the overall compilation flow.

\subsection{Compiler Optimization Using Reinforcement Learning}
After a framework as described above is set up, 
it can be used as the basis for learning the best sequence of actions (i.e., compilation passes) depending on characteristics of the initial circuit.
Motivated by its success in classical compiler optimization~\cite{reinforcement1,reinforcement2,reinforcement3}, we propose to apply reinforcement learning for this task.
RL-methods aim to learn an \emph{action policy} that describes which action to apply 
based on certain \emph{observations} on the current state such that some cumulative \emph{reward function} is maximized.
In contrast to supervised machine learning methods, no training data must be provided.
Instead, only an environment describing the underlying MDP (i.e., the set of states and corresponding actions), the features used as observations, and a corresponding reward function needs to be provided.
Based on the rewards for certain actions in certain states, the action policy improves through the training process and learns how to maximize the expected cumulative reward.

In the proposed framework, 
observations are realized by extracting certain features from the quantum circuits in each state.
In addition, a sparse reward function is used that reflects whether a final state has been reached and, then, quantifies the \enquote{quality} of the resulting circuit, e.g., in terms of the resulting gate count, circuit depth, or expected fidelity.
This flexibility in the target metric allows one to design specialized compilers for dedicated use cases.
After successful training, the learned action policy can be used as an \emph{optimized compiler} for any quantum circuit.

\subsection{Related Work}\label{sec:related}

Machine learning has been successfully applied to quantum circuit design/compilation in the past.
In~\cite{ml_alexandru}, the authors propose a machine learning based layout and routing scheme and claim to be \mbox{on-par} with \mbox{state-of-the-art} methods.
Another possible application is given in~\cite{wang2022quest}, where the graph structure of quantum circuits is used to predict their expected fidelities for certain devices considering their respective noise characteristics.
An idea related to the one proposed in this work, which follows a more \mbox{coarse-grained} approach, has been demonstrated in~\cite{quetschlich2022mqtpredictor}.
There, a supervised machine learning model is used to predict good combinations of devices and existing compilation flows.
In~\cite{rloptgoogle2021}, RL has been applied to learn a strategy for applying individual gate transformation rules to optimize quantum circuits.

While these works either optimize one specific type of action or predict compilation options on a high abstraction level within one quantum SDK, the framework proposed in this work results in \mbox{fine-grained} steps that allow one to stitch together methods from multiple SDKs with a customizable optimization objective into a \mbox{fully-fledged} compilation flow.

\section{Feasibility Study}\label{sec:exp}

In the following, the feasibility of the proposed framework is demonstrated.
To this end, we first describe the particular instantiation that has been developed as part of this work.
Then, the performance of the resulting optimized compiler is evaluated and compared against the \mbox{state-of-the-art} compilers available in IBM's Qiskit~\cite{qiskit} and Quantinuum's TKET~\cite{tket}.
All developments and results are made publicly available on GitHub (\url{https://github.com/cda-tum/MQTPredictor}) as part of the Munich Quantum Toolkit (MQT).

\vspace{-1mm}
\subsection{Instantiation}

The basis of the framework is formed by the open-source \emph{OpenAI Gym}~\cite{openaigym} library, which is used to set up an environment for the MDP described in \autoref{sec:proposed} (as illustrated in \autoref{fig:mdp}).
In this regard, the main endeavor is the selection and integration of all compilation passes that should be considered as actions in the learning process.
As modeled in \autoref{fig:mdp}, we distinguish five kinds of actions: Quantum platform selection, Quantum device selection, Synthesis, Mapping, and Optmization.
Due to the unified action interface of the proposed framework described earlier, methods from more than one SDK or toolkit can be incorporated.
For each kind of action, suitable representatives from IBM's Qiskit (version $0.39.2$) and/or Quantinuum's TKET (version $1.8.1$) have been chosen---leading to the following list of available actions.
\begin{itemize}
	\item \emph{Platforms} and corresponding \emph{Devices}:
	\begin{itemize}
		\item IBM (superconducting): \emph{ibmq\_montreal} (27 qubits) and \emph{ibmq\_washington} (127 qubits)
		\item Rigetti (superconducting): \emph{Aspen-M-2} (80 qubits)
		\item IonQ (trapped-ions): \emph{Harmony} (11 qubits)
		\item OQC (superconducting): \emph{Lucy} (8 qubits)
	\end{itemize}
	\item \emph{Synthesis}: Qiskit's \emph{BasisTranslator}
	\item \emph{Mapping}: Any combination of the following methods
	\begin{itemize}
		\item Layout:
		\begin{itemize}
			\item Qiskit's \emph{TrivialLayout}
			\item Qiskit's \emph{DenseLayout}
			\item Qiskit's \emph{SabreLayout}
		\end{itemize}
		\item Routing:
		\begin{itemize}
			\item Qiskit's \emph{BasicSwap}
			\item Qiskit's \emph{StochasticSwap}
			\item Qiskit's \emph{SabreSwap}
			\item TKET's \emph{RoutingPass}
		\end{itemize}
	\end{itemize}
	\item \emph{Optimization}:
	\begin{itemize}
		\item Qiskit's \emph{Optimize1qGatesDecomposition} 
		\item Qiskit's \emph{CXCancellation} 
		\item Qiskit's \emph{CommutativeCancellation} 
		\item Qiskit's \emph{CommutativeInverseCancellation} 
		\item Qiskit's \emph{RemoveDiagonalGatesBeforeMeasure} 
		\item Qiskit's \emph{InverseCancellation} 
		\item Qiskit's \emph{OptimizeCliffords} 
		\item Qiskit's \emph{Collect2qBlocks} + \emph{ConsolidateBlocks} 
		\item TKET's \emph{PeepholeOptimise2Q}
		\item TKET's \emph{CliffordSimp}
		\item TKET's \emph{FullPeepholeOptimise}
		\item TKET's \emph{RemoveRedundancies}
	\end{itemize}
\end{itemize}

Inspired by the feature selection and demonstrated relevance in~\cite{quetschlich2022mqtpredictor}, the observations used to guide the reinforcement learning agent are based on seven features---namely the \emph{number of qubits}, the \emph{depth} of the circuit, and the five composite features of \emph{program communication}, \emph{\mbox{critical-depth}}, \emph{\mbox{entanglement-ratio}}, \emph{parallelism}, and \emph{liveness} originally proposed in~\cite{supermarq}.
For the learning itself, three different reward functions have been implemented:
\begin{enumerate}
	\item \emph{Expected fidelity}: provides an estimate on the reliability of the circuit based on characteristics of the chosen device. A fidelity of $1$ indicates an error-free result, while a fidelity of $0$ indicates that the result is completely wrong.
\item \emph{Critical Depth}: describes how many \mbox{two-qubit} gates in a quantum circuit lie along the critical path and contribute to the overall circuit depth. A critical depth close to $1$ indicates a highly sequential quantum circuit. Since, a lower value is desirable, the corresponding reward function is defined as $1-\text{critical depth}$.
\item \emph{Combination}: adds up both mentioned reward functions (divided by two to result in a value between $0$ and $1$).
\end{enumerate}

Finally, the \emph{Proximal Policy Optimization (PPO)} algorithm~\cite{ppo_algo} provided by \emph{\mbox{Stable-Baselines3}}~\cite{stable-baselines3} is used for the reinforcement learning.

\vspace{-2mm}
\subsection{Results}
\vspace{-1mm}
\begin{figure*}[t]
     \begin{subfigure}[b]{0.32\textwidth}
         \centering
         \includegraphics[width=\textwidth]{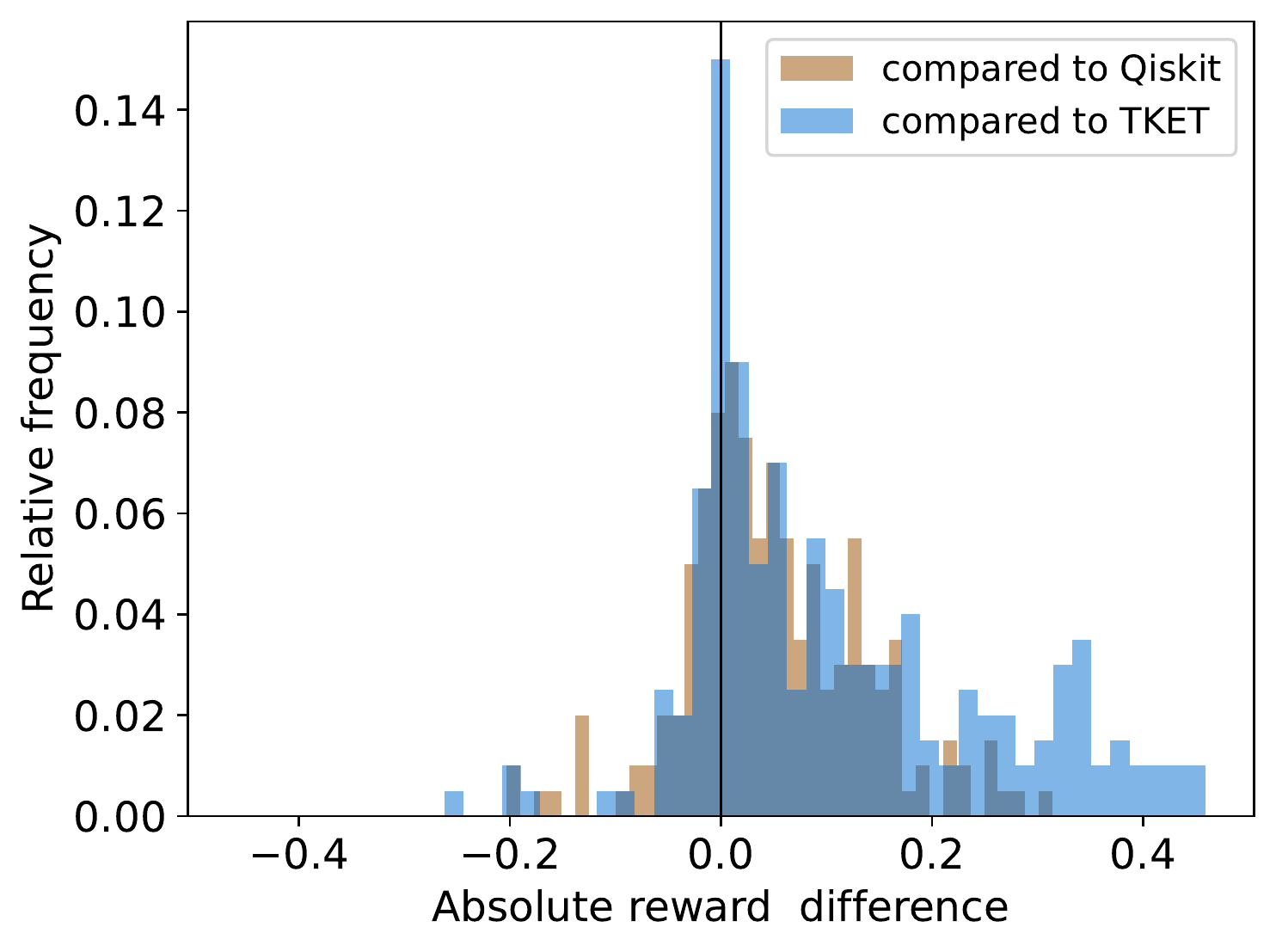}
         \caption{Fidelity}
         \label{fig:fid_hist}
     \end{subfigure}
     \hfill
     \begin{subfigure}[b]{0.32\textwidth}
         \centering
         \includegraphics[width=\textwidth]{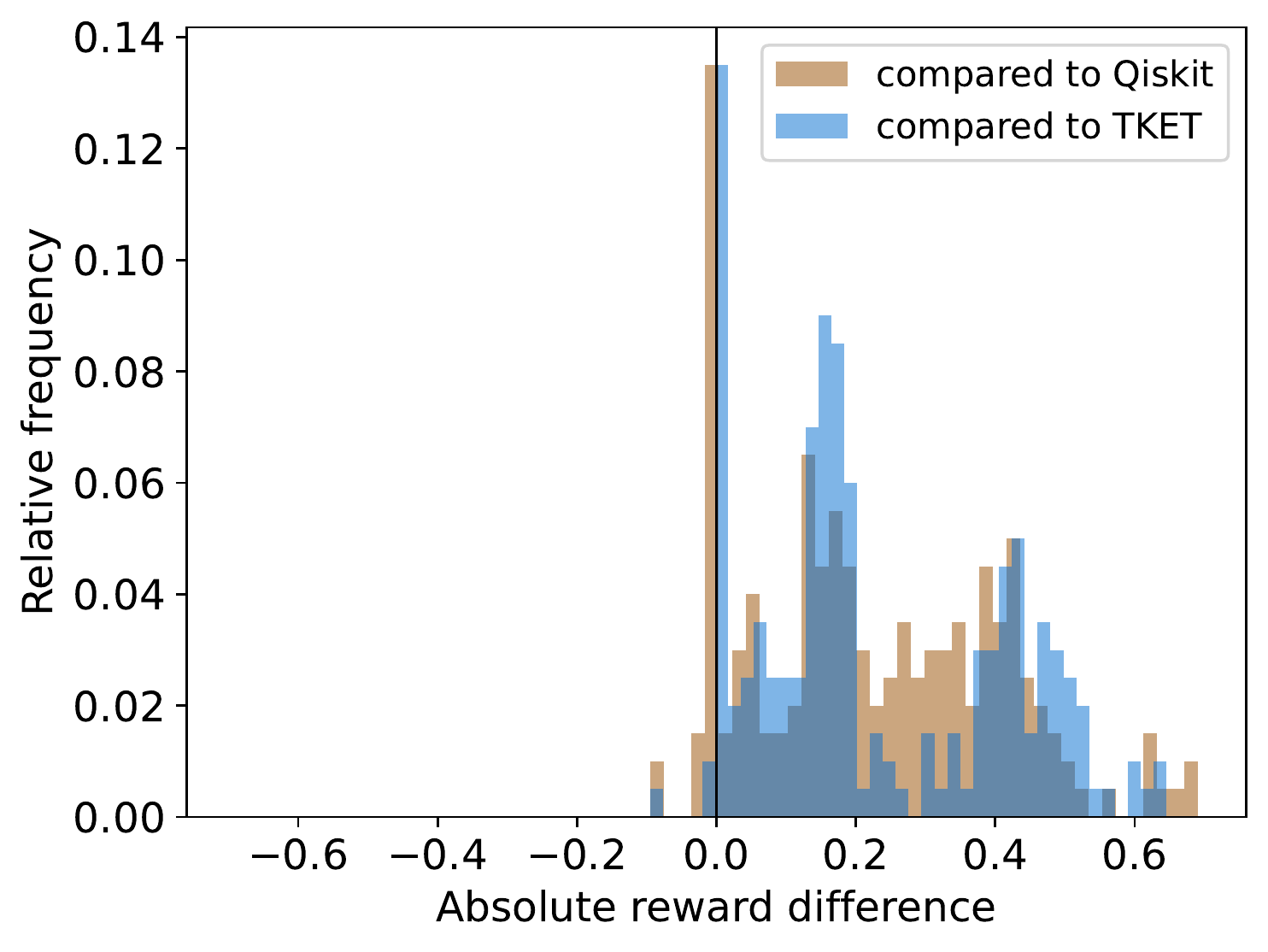}
         \caption{Critical Depth}
         \label{fig:depth_hist}
     \end{subfigure}
     \hfill
     \begin{subfigure}[b]{0.32\textwidth}
         \centering
         \includegraphics[width=\textwidth]{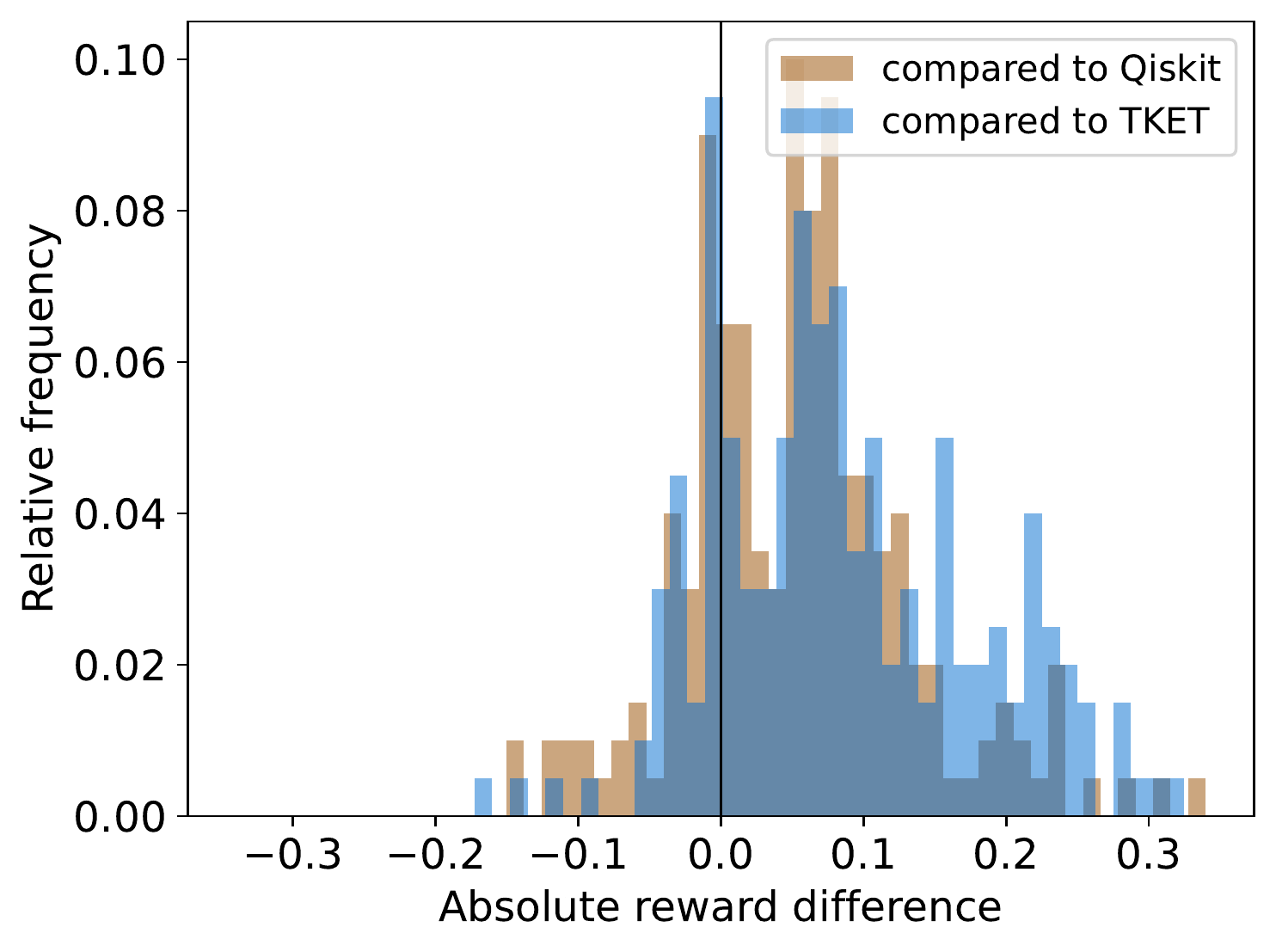}
         \caption{Combination}
         \label{fig:mix_hist}
     \end{subfigure}

    \begin{subfigure}[b]{0.32\textwidth}
         \centering
         \includegraphics[width=\textwidth]{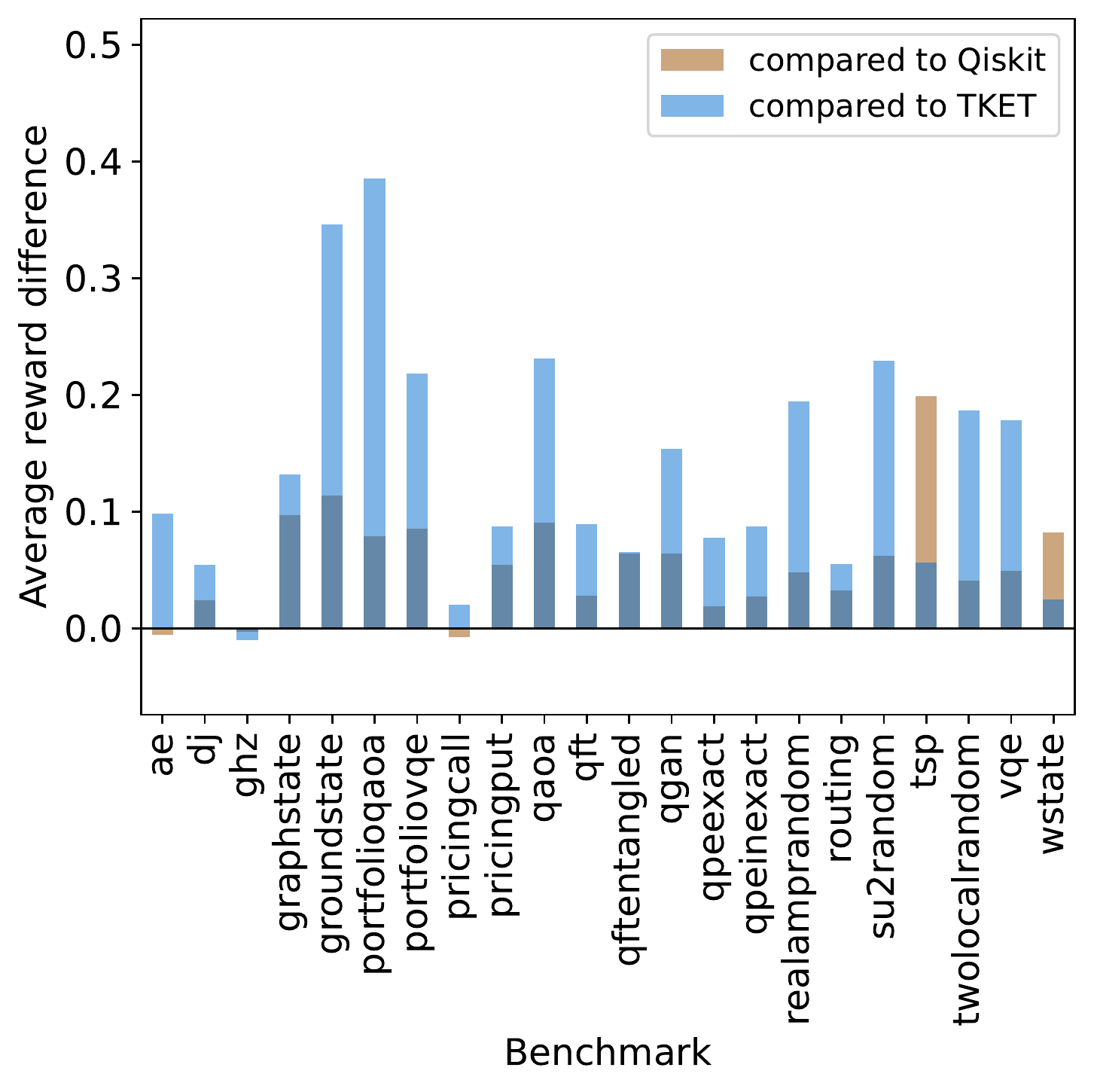}
         \caption{Fidelity}
         \label{fig:fid_bench}
     \end{subfigure}
     \hfill
     \begin{subfigure}[b]{0.32\textwidth}
         \centering
         \includegraphics[width=\textwidth]{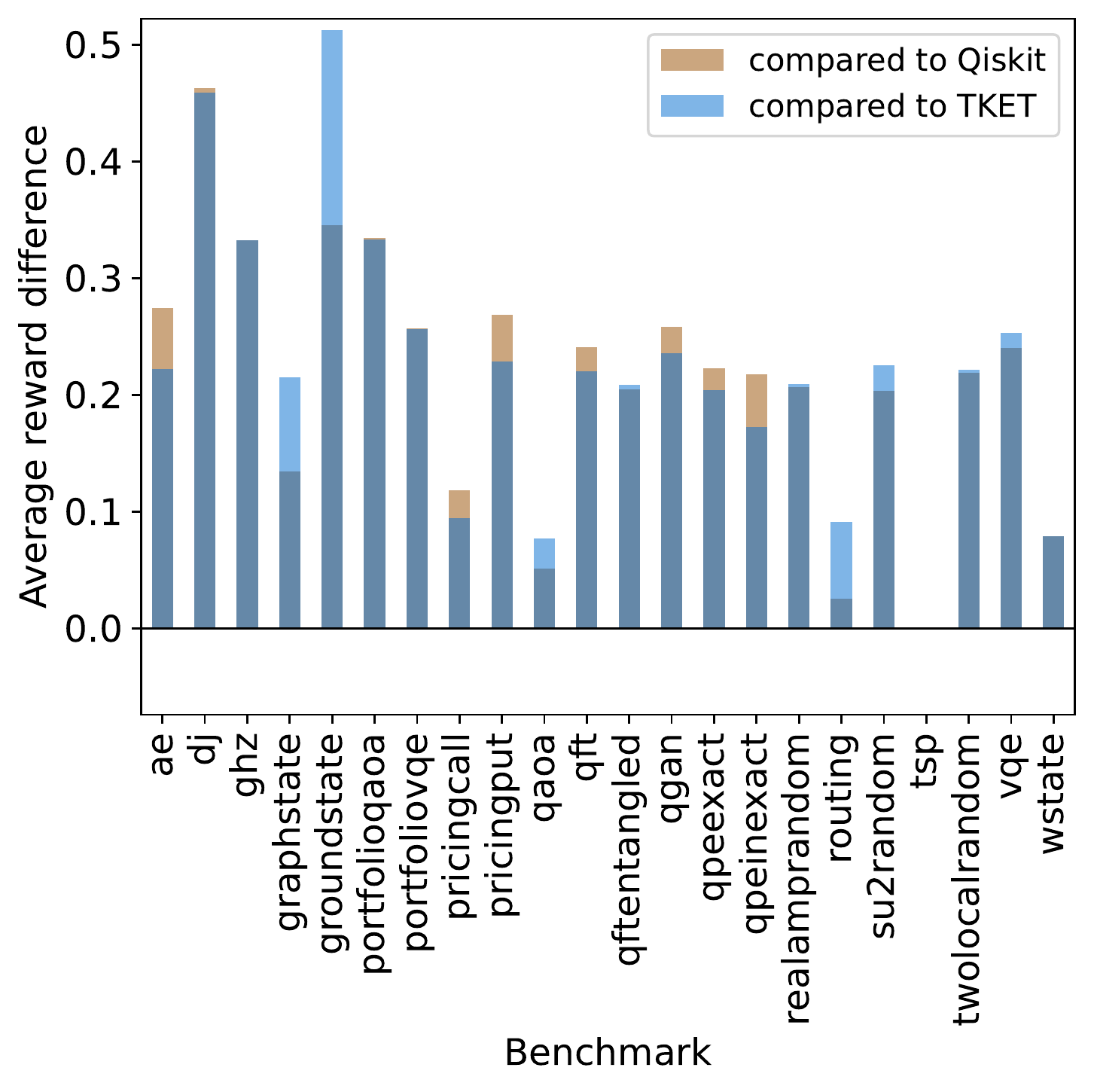}
         \caption{Critical Depth}
         \label{fig:depth_bench}
     \end{subfigure}
     \hfill
     \begin{subfigure}[b]{0.32\textwidth}
         \centering
         \includegraphics[width=\textwidth]{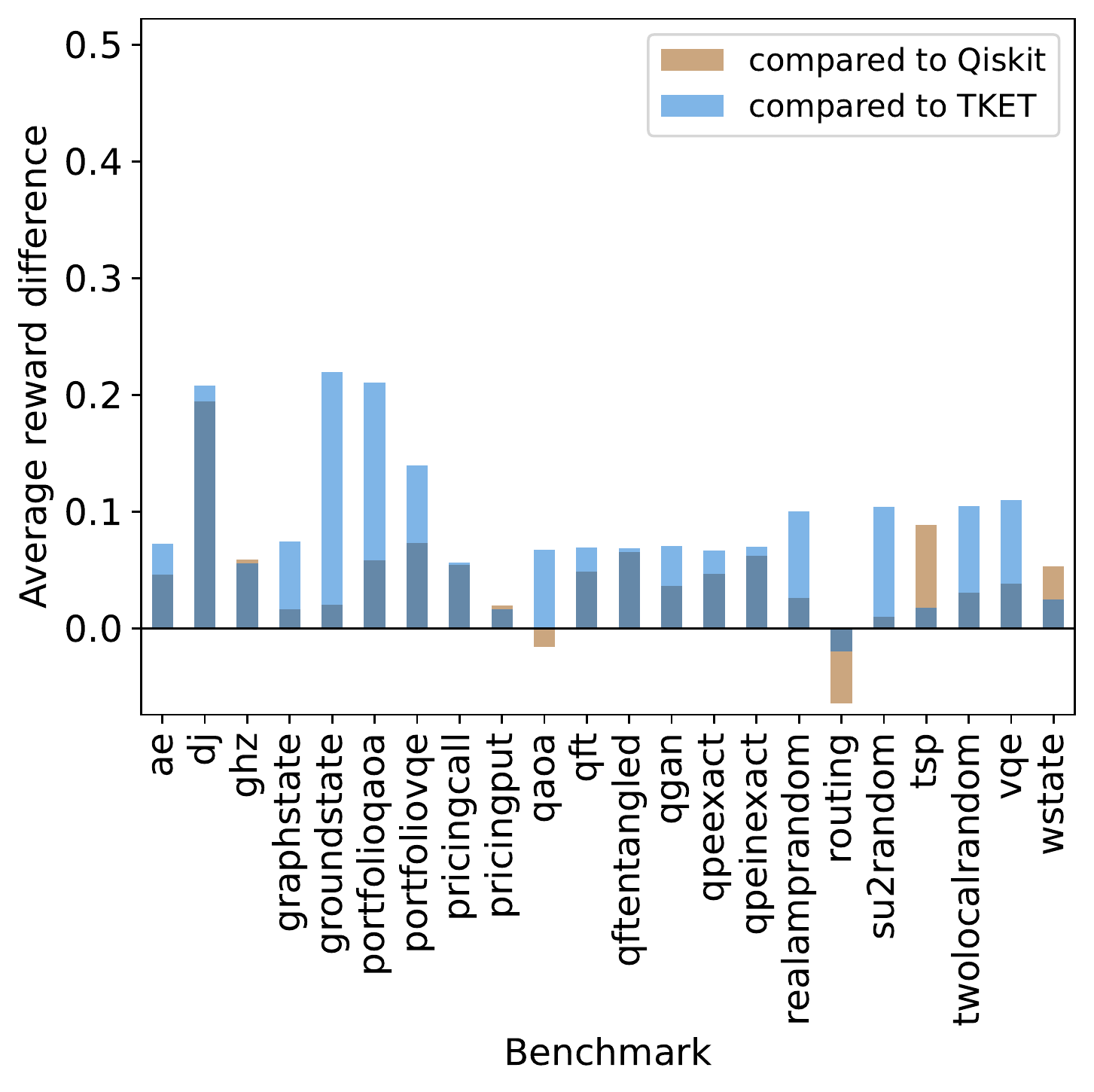}
         \caption{Combination}
         \label{fig:mix_bench}
     \end{subfigure}
     
     \vspace{-2mm}
        \caption{Experimental evaluation.}
         \label{fig:exp_eval}
         \vspace{-7mm}
\end{figure*}

To evaluate the performance of the framework instantiation described above, three models---one for each reward function---were trained on a selection of $200$ circuits ranging from $2$ to $20$ qubits taken from the \emph{\mbox{target-independent}} level of the MQT Bench library~\cite{quetschlich2022mqtbench}.
In total, $100,000$ time steps were executed during the training for each reward function with runtimes of the order of hours. 
Then, the trained models were evaluated on the same circuits and the resulting quality in terms of the respective reward function was measured.
To put this performance into perspective, all circuits are additionally compiled to the \emph{ibmq\_washington} device using Qiskit and TKET with the highest optimization level ($O3$ for Qiskit and $O2$ for TKET).

The results are summarized in \autoref{fig:exp_eval}.
To this end, \autoref{fig:fid_hist}, \autoref{fig:depth_hist}, and \autoref{fig:mix_hist}, respectively show the distribution of \emph{absolute} reward differences of the proposed approach compared to Qiskit and TKET with respect to the three quality metrics.
An $x$-value larger than zero indicates a positive improvement, e.g., a value of $0.2$ for the fidelity means that the proposed approach managed to boost the fidelity by $20\%$.
In an ideal scenario, all results would accumulate on the right-side of the zero-line.

The results demonstrate that, in the vast majority of cases, this is the case, i.e., the proposed method produces circuits of equal or better quality.
In particular, the proposed method outperforms the Qiskit/TKET compiler in $73\%$/$80\%$, $84\%$/$86\%$, and $75\%$/$78.5\%$ of cases regarding expected fidelity, critical depth, and their combination.

The advantage of the proposed method becomes even clearer when looking at the average reward difference per benchmark algorithm (subsuming several instances with varying amounts of qubits) as shown in \autoref{fig:fid_bench}, \autoref{fig:depth_bench}, and \autoref{fig:mix_bench} respectively.
There, $y$-values higher than zero indicate a net gain in quality.
Again, in the vast majority of cases, a substantial improvement can be observed.
Compared to Qiskit/TKET, an average \emph{absolute} increase in fidelity, critical depth, and their combination of $4.9\%$/$10.7\%$, $22.6\%$/$22.8\%$, and $5.5\%$/$8.5\%$ is achieved.

To further underline the proposed framework's ability to adapt to different target metrics, we additionally calculated the average reward for each metric by all three trained models---not only the one trained for the respective metric.
The resulting numbers are denoted in \autoref{tab:comparison}.
They confirm that the model trained for a particular purpose indeed results in the compiled circuits with the highest rewards for all three cases.

\begin{table}[t]
\caption{Comparison of reinforcement learning models.}
\label{tab:comparison}
\vspace{-3mm}
\centering
\resizebox{.99\linewidth}{!}{
\begin{tabular}{l|ccc}
& \multicolumn{3}{c}{Average result for...} \\ 
Model trained for... & Fidelity  & Critical depth & Combination\\ 
\hline 
Fidelity & \textbf{0.48}  & 0.27 & 0.37 \\ 
Critical depth & 0.18   & \textbf{0.47} & 0.33\\ 
Combination & 0.45 & 0.33 & \textbf{0.39}  \\ 
\end{tabular} 
}
\vspace{-7mm}
\end{table}

\vspace{-3mm}
\section{Conclusions}\label{sec:conclusions}
\vspace{-2mm}
Quantum circuit compilation flows are becoming more and more complex.
This leads to a similar situation observed in classical compilation where there are so many options and it is unclear which options to choose and in which order to apply them.
In this work, we proposed a framework that builds on the decades of research on classical compiler optimization to overcome these obstacles.
To this end, the quantum circuit compilation flow was modeled as a Markov Decision Process with distinct constraints for each state and a unified interface for all actions.
Based on that, reinforcement learning was applied to derive a complete compilation flow out of actions from multiple SDKs and toolkits.
Experimental evaluations on the developed framework and three different optimization objectives demonstrate the advantage of the proposed method over Qiskit’s and TKET’s most optimized compilation sequences and the general feasibility of automatically creating compilations flows targeting a customizable objective.
In $73\%$, $84\%$, $75\%$ of the cases, the trained models outperformed Qiskit/TKET with an average \emph{absolute} improvement of $4.9\%$/$10.7\%$, $22.6\%$/$22.8\%$, and $5.5\%$/$8.5\%$ regarding expected fidelity, critical depth, and their combination respectively.
These results clearly demonstrate that adapting techniques from the classical realm can, indeed, yield substantial improvements in the quantum realm.
By making the developed framework publicly available as \mbox{open-source}, we hope to lay the foundation for further exploration of this important area of research.

\vspace{-3mm}
\section*{Acknowledgments}
\vspace{-2mm}
This work received funding from the European Research Council (ERC) under the European Union’s Horizon 2020 research and innovation program (grant agreement No. 101001318), was part of the Munich Quantum Valley, which is supported by the Bavarian state government with funds from the Hightech Agenda Bayern Plus, and has been supported by the BMWK on the basis of a decision by the German Bundestag through project QuaST, as well as by the BMK, BMDW, and the State of Upper Austria in the frame of the COMET program (managed by the FFG).

\printbibliography

\end{document}